# Earthquake damage patterns resolve complex rupture processes

**Authors:** Yann Klinger[1]*, Kurama Okubo[1,2], Amaury Vallage[1], Johann Champenois[1,3], Arthur Delorme[1], Esteban Rougier[4], Zhou Lei[4], Earl E. Knight[4], Antonio Munjiza[5,] Stephane Baize[3,] Robert Langridge[6] and Harsha S. Bhat[2]

**Affiliations :**
[1]Institut de Physique du Globe de Paris, Sorbonne Paris Cité, Université Paris Diderot, CNRS, Paris, France.
[2]Laboratoire de Géologie, École Normale Supérieure/CNRS UMR8538, PSL Research University, Paris 75005, France
[3]Seismic Hazard Division, Institut de Radioprotection et de Sûreté Nucléaire, Fontenay-aux-Roses, France
[4]EES-17 – Earth and Environmental Sciences Division, Los Alamos National Laboratory, NM, USA
[5]FGAG– University of Split, 21000 Split, Croatia.
[6]GNS Science, P.O. Box 30-368, Lower Hutt, 5040, New Zealand.

*Corresponding author. Email: klinger@ipgp.fr.

**Large continental earthquakes activate multiple faults in a complex fault system, dynamically inducing co-seismic damage around them. The 2016 Mw 7.8 Kaikoura earthquake in the northern South Island of New Zealand has been reported as one of the most complex continental earthquakes ever documented[1], which resulted in a distinctive on and off-fault deformation pattern. Previous geophysical studies confirm that the rupture globally propagated northward from epicenter. However, the exact rupture-propagation path is still not well understood because of the geometrical complexity, partly at sea, and the possibility of a blind thrust. Here we use a combination of state-of-the-art observation of surface deformation, provided by optical image correlation, and first principle physics-based numerical modeling to determine the most likely rupture path. We quantify in detail the observed horizontal co-seismic deformation and identify**

**specific off-fault damage zones in the area of the triple junction between the Jordan, the Kekerengu and the Papatea fault segments. We also model dynamic rupture propagation, including the activation of off-fault damage, for two alternative rupture scenarios through the fault triple junction. Comparing our observations with the results from the above two modeled scenarios we show that only one of the scenarios best explains both the on and off-fault deformation fields. Our results provide a unique insight into the rupture pathway, by observing, and modeling, both on and off-fault deformation. We propose this combined approach here to narrow down the possible rupture scenarios for large continental earthquakes accompanied by co-seismic off-fault damage. Thus combining observations and numerical modeling of both on and off-fault deformation fields opens avenues for understanding complex rupture patterns, including those of past earthquakes whose off-fault deformation zones are still preserved.**

Large crustal earthquakes result from ruptures that dynamically propagate through a complex network of faults, whose temporal sequence of failure is not always clear[1-3]. Associated secondary faulting and co-seismic off-fault damage suggest that a significant part of on and off-fault deformation patterns are due to state of traction, fault geometry and directivity of the rupture[4-6], in addition to some geological structural inheritance[7]. At ground surface this off-fault damage zone can be hundreds of meter wide[8,9], while it becomes narrower at depth[10]. The combined length of surface ruptures associated with the 13$^{th}$ November 2016 $M_w$ 7.8 Kaikoura earthquake in New Zealand (Fig. 1) reaches 180 km, distributed over more than 15 distinct fault segments[1,11]. Although a blind low-angle thrust might have been activated[12], the right-lateral strike-slip faults oriented NE-SW, such as the Jordan and the Kekerengu faults, dominate surface ruptures[11,13,14]. The 15 km-long NNW-SSE Papatea fault segment, however, is characterized by left-lateral motion of up to ~6 m and by vertical throw reaching 10 m[15].

Surface-rupture observations show that the northern tip of the Papatea fault does not connect to the Jordan - Kekerengu fault system[15]. All geophysical studies agree that the rupture propagated northward from the epicenter[1,16-18]. However, because of the geometrical complexity, partly at sea, and the possibility of a blind thrust[12], the exact rupture-propagation path remains elusive; in particular the way the rupture propagated through the Papatea – Jordan - Kekerengu triple junction remains unknown and the Papatea fault is generally ignored in rupture models.

**High-resolution optical satellite image correlation**

Using optical satellite images bracketing the date of the Kaikoura earthquake, we measured amplitude and direction of the horizontal displacement field in the triple junction area (Fig. 2). SPOT6 images (resolution 1.8m) pre-dating the earthquake were correlated with Pleiades images (resolution 0.5 m), acquired between December 2016 and March 2017, using MicMac[19] (*see Method section*). Although our measurements might include post-seismic deformation, it should be less than 10% of the co-seismic deformation[12] and should not affect significantly our observations. Thus, ground resolution of our displacement field is 1.8 m, with a displacement detection threshold of about 20 cm[19].

Fig. 2 shows the amplitude of the horizontal displacement at the triple junction. Systematic swath profiles every 90 m across different fault segments allow to establish a detailed slip distribution for that part of the Kaikoura rupture (Figs. S1, S3). Along the Jordan and the Kekerengu faults, 8 km-long swath profiles J1, J2, K1 and K2 (Figs. 2, S1) show displacement parallel to the fault, where the full strike-slip deformation is highly localized in a band only a few tens of meters wide. Along the Kekerengu fault, we measured a maximum right-lateral co-seismic displacement of about 11 m (Fig. S3), in good agreement with the direct field-offset measurements[15]. This displacement field reveals that the pattern of deformation along the Papatea fault differs significantly from patterns along Kekerengu and Jordan faults. Along the Papatea fault, swath profiles P1 to P5 show that the gradient of horizontal deformation is not sharp everywhere (Figs. 2, S1). Instead, at both extremities, the displacement gradient is less sharp, which is interpreted as distributed deformation across a damaged fault zone. Thus, the total 6m left-lateral displacement measured along P1 is distributed over a width of 2 kilometers, which is consistent with field[15] and Lidar mapping[20] that documented several parallel fault strands at the coast. Actual fault scarps in the deformation zone are also visible on the profile. Along profiles P2 the deformation zone becomes narrower and asymmetric relative to the position of the fault, with most of the distributed deformation located south of Papatea. Profiles P3, P4 and P5, located north of the major bend of the Papatea fault, show that the damage zone becomes wider again, to eventually include the entire triangular zone bounded by the Papatea fault, the Kekerengu fault, and to the North-East, by the short Waiautoa fault (Fig. 2).

**Earthquake rupture modeling and off-fault damage pattern**

To elucidate the rupture scenario that best explains the observed displacement field, we consider two hypothetical cases (Fig. 1). In the first scenario, the rupture propagated northward from the epicenter to reach the northern tip of the Hundalee fault and continued northward, offshore, until it would trigger slip on the Papatea fault. This scenario is consistent with observed co-seismic uplift of the Kaikoura peninsula[20,21], observation of submarine surface ruptures along the Point Kean Fault [11], and numerical models of the entire rupture[22]. In the second scenario, the rupture propagated northward from the epicenter to reach the northern tip of the Hundalee fault and jumped about 20 km to the NW to dynamically trigger rupture along the Jordan fault.

We model the two scenarios using a 2D continuum-discontinuum model (*see Method section*) that allows for dynamic rupture propagation on prescribed faults (Fig. 2, S6) and for spontaneous activation of off-fault fracture damage. Numerical simulation of each scenario led to a distinctive pattern of rupture sequence and off-fault damage.

In the first scenario, the rupture first propagates northward along Papatea and jumps on to the Jordan-Kekerengu fault system. This rupture then propagates bilaterally from the junction (Fig. 3a-e, MS1). While the rupture is propagating along the Papatea fault, significant damage occurs on the southern side, around the main kink of the fault (Fig. 3a-e). A major zone of damage also develops in the triangular zone between Kekerengu, Papatea and Waiautoa faults. No significant damage, however, occurs along the Jordan fault (Fig. 3f). This rupture scenario is in good agreement with observations and other numerical models[22]. In addition to the rupture, we have also managed to capture the off-fault displacement field, due to damage (Figs. 3g, 3h). We non-dimensionalize the spatial distribution and the amplitude of displacement, for comparison with data, as our aim is to capture the broad features of the displacement field and not the specifics of the slip distribution. Regardless, we show very clearly that off-fault damage has to be taken into account to explain the rupture path and the on and off- fault displacement fields which cannot be recovered by utilizing purely elastic models (dashed lines, Fig. 3g, h).

In the second scenario, the rupture jumps from Jordan to Papatea (Fig. S6c, MS2) and is immediately arrested due to significant off-fault damage (Fig. S6d, e). The southern part of the Papatea fault does not rupture while the rupture continues on Kekerengu. The prominent damage is mostly off Kekerengu (Fig. S6, f) and very little off Papatea. This is neither in agreement with the observed displacement field (Fig. S6g, h) nor with the observed surface rupture (Fig. 4).

Comparison of the swath profiles through the horizontal displacement field with damage patterns resulting from each scenario (Figs. 3g, h, S6g, h) shows that the first scenario is more consistent with observations: both in observations and models, patches of damage are localized at the kink in the hanging wall of Papatea, and in the triangular zone located NW of Waiautoa. Another discriminant is the absence of damage NW of Papatea fault in observations and in the numerical model (Fig. S6).

**Kaikoura earthquake rupture path**

In summary, as seen in Fig. 4, both the spatial pattern of damage and the field observations, when confronted with the two modeled rupture scenarios, suggest that the rupture did propagate along the Papatea fault, from the coast to the triple junction area where it triggered a bi-lateral rupture on the Jordan Thrust – Kekerengu fault system. In addition, detailed field observation of the surface ruptures along the Papatea and the Waiautoa faults reveals that in several places, secondary ruptures systematically branch off from the main fault scarps toward NW, in a pattern compatible with a left-lateral strike-slip rupture propagating toward the northeast[6] (Fig. 4). This observation also supports the first rupture scenario. Although it is at the limit of the resolution of seismological data available, the seismic source studies that are focused on the second part of the Kaikoura rupture are also compatible with this scenario[13,23].

Hence, although the $M_w$ 7.8 Kaikoura earthquake has been deemed one of the most complex continental earthquake ruptures ever documented because of the very large number of fault sections activated, the general rupture mechanism might actually be simple. From the epicenter, the rupture propagated northward, navigating local geometrical complexities, extended off-shore along the Hundalee fault and then along the Point Kean fault. Eventually it dynamically triggered a rupture along the Papatea fault, located at a maximal distance of 12 km, although it might be closer off-shore. The rupture then propagated northward along Papatea and eventually triggered a bi-lateral rupture along the Jordan – Kekerengu fault system. The Papatea block acted as a large-scale compressional jog, which is consistent with the large documented uplift[1,24].

At first glance surface ruptures might appear very complex during large continental earthquakes such as the Kaikoura earthquake. This complexity, however, can be resolved and the rupture follows a rather simple structural path. *This rupture path can be discerned by measuring, and modeling, both on and off-fault damage patterns*. Earthquake simulators used in seismic hazard assessments for complex fault systems, such as for the Southern California fault system[25], can generate myriads of very large and complex fault ruptures. Providing critical keys, like off-fault damage patterns, to decipher this complexity might help narrow down a subset of most probable scenarios along complex fault networks.

**Methods**

**Image correlation processing:**

To measure the horizontal displacements associated with the 2016 Kaikoura event, we have correlated optical satellite images acquired before and after the earthquake. The correlation processing has been conducted by using the open-source software package MicMac[19].
Two correlation maps with three different types of images have been computed:

Low-resolution correlation: Two Sentinel-2 images have been used, which have been acquired respectively on April 9th, 2016 and on December 15th, 2016. The Sentinel-2 images are multispectral images with pixel resolution varying between 10m and 60m, depending on the wavelength. Here, we have used the 4 bands with a pixel-resolution of 10m: Red, green, blue and near infrared. These images are ortho-rectified by the image provider (European Space Agency) and can be correlated without any specific pre-processing. For each band, each pair of images (pre- and post-earthquake images) has been processed independently to obtain 4 displacement maps for each component of the displacement, the North-South component and the East-West component. To improve the signal-to-noise ratio, for each component of displacement the 4 maps have been merged, based on the median value for each quadruplet of pixels. The result of the Sentinel-2 correlation is presented in Fig. S2. The displacement values can be crosschecked against GPS and static displacement recorded by local strong-motion instruments. The far-field displacement is set to zero. Our results compare well with previously published horizontal-displacement fields computed from space geodesy and GPS measurements[1,12-14,17,24].

High-resolution correlation: To image details of the deformation in the close vicinity of the surface rupture, we have performed correlation of metric-scale images. For the images before the Kaikoura earthquake, we used a stereo-pair of images acquired by the satellite SPOT6 on May 18th, 2014. For the images acquired after the earthquake, we use a combination of several tri-stereo images acquired by the satellite Pleiades (operated by the French space agency CNES) between December 23rd 2016 and March 18th 2017. To ensure the best processing of the different multiplets of Pleiades images, they have been processed separately and only the final correlation maps were merged. The area of interest has been limited to the triple junction area between the Jordan thrust-Kekerengu fault system and the Papatea fault system.

A pre-earthquake digital elevation model (DEM) was computed at the resolution 1.8m from the SPOT6 images, and a post-earthquake DEM was computed at the resolution 0.5m from the Pleiades. Both DEM were computed using the Micmac package[19]. These DEMs were used to ortho-rectify the different sets of images in order to be able to correlate them. Because the two sets of images are not originating from the same sensor and they have different native resolution, the Pleiades images had to be resampled at 1.8m to be consistent with the SPOT6 images. This resampling has been done using the open-access GDAL library. After ortho-rectification, because it has been done independently for the images SPOT6 and Pleiades, a final adjustment (~10m) has been done by applying a rigid translation to the Pleiades images, based on ground control points (GCP) identified on both image datasets and located far from major surface ruptures. Then, the pre- and post-earthquake images have been correlated to compute the horizontal-displacement field at the resolution of 1.8m. To ensure that no long-wavelength noise was contaminating the high-resolution displacement field, which could be due to imperfect ortho-rectification or correction for satellite attitude, the result of the high-resolution correlation has been compared to the Sentinel-2 correlation that were validated with external data (GPS and ground-motion data, see Fig. S2) and the difference was corrected by removing a linear ramp estimated through a root-mean-square best fit.

**Modeling dynamic earthquake rupture with co-seismic off-fault damage by continuum-discontinuum approach:**

We use a continuum-discontinuum based scheme, the combined finite-discrete element method (FDEM), to achieve both high-numerical accuracy of rupture propagation, seismic wave radiation and to model the activation of new cracks, in both tensile and shear, in the off-fault medium. We used the FDEM based software tool, Hybrid Optimization Software Suite – Educational Version (HOSSedu) developed by Los Alamos National Laboratory, for all simulations in this study. We firstly trace a part of the entire fault system close to the triple junction from observations as shown in Fig. S4 (a). We then discretize the domain by using an unstructured triangular mesh around the prescribed faults. The mesh size is adaptively controlled to be finer close to the fault to optimize trade-off between the numerical accuracy and computational cost. We then define the initial stress state $\sigma_{ij}$ uniformly in the medium. The angle of $\sigma_1$, the maximum principal compressive stress, is limited to be compatible with the sense of slip on the faults and hence is restricted to be in a range of 105°-115° with respect to North. The direction of the maximum principal stress was chosen to be N107° to be compatible with both the rupture scenarios and regional focal mechanisms[26]. It is assumed that the material around the faults has been previously damaged (i.e. weakened) and therefore is less competent that the rest of the material in the model. The areas of weakened material are highlighted in yellow in Figure S4a. The introduction of this weakened material area will also restrict unrealistic crack propagation at the edge of fault generated by the fact that a relatively simple friction model (friction slip weakening law) was used in this case. The FDEM allows for tensile, shear and mixed-mode crack represented as the break of cohesion at the boundary of the finite elements. In other words, each boundary of a finite element is a potential failure plane. To avoid numerical bias in the orientation of cracks, the orientations of the potential failure planes are kept isotropic as shown in Fig. S4 (b). Figures. S4 (c) and (d) describe the failure criteria in FDEM. Two types of interaction forces, cohesion and contact/friction, are operating at each boundary of the finite elements. The method is explicit in terms of time integration, so the governing equations are solved on an element-by-element basis. The evolution of the cohesive and frictional forces at the interfaces as a function of the relative displacements are shown in Fig. S4 (c) and (d). The opening and shear displacements, $\delta_I$ and $\delta_{II}$, are used to derive the cohesive forces at each time step. The first portion of the curves describing the cohesion as a function of the displacements represents a non-linear elastic loading part, which occurs over a very small range of relative displacements between any two boundaries of finite elements. Within this range of deformation at the boundary of the finite elements, it is ensured that the entire medium behaves purely elastically since the finite element deforms as purely an elastic medium satisfying the elastic constitutive law. When the traction on the boundary of an element reaches the peak strength, then damage starts to accumulate and the cohesive strength of the interface starts to decrease linearly up to a point where it is eventually totally broken. The dissipated energy is represented by the area of triangles highlighted with slanted lines, see Fig. S4 (c). The friction curve also features an elastic loading portion, followed by the conventional linear slip-weakening law. We resolve both cohesion and friction at the interfaces of the finite elements located on the off-fault medium and only friction at the interfaces of the finite elements located along the prescribed faults; this implies that the off-fault medium is considered to be intact at the beginning of the dynamic rupture modelling. Since the fracture energy in shear is proportional to the amount of slip[27], the friction parameters differ between the main prescribed fault and the off-fault medium. The amount of slip is, on average, one or two orders of magnitude higher along the main fault than in the off-fault medium. When the cohesion between the finite elements starts to break, we visually plot the dynamically generated cracks as highlighted in red in Fig. S4 (e). The values of parameters used in our modelling are listed in Table S1. The main algorithmic solutions utilized within HOSSedu are described in detail in a series of monographs[28].

In general, the material constants, the initial stress state and the frictional properties play a key role in the dynamic earthquake rupture processes. We employed a homogeneous medium, which has common material properties similar with granite. We then determine the peak cohesive strength for cohesion based on the closeness to failure (CF), which is defined as the ratio of the radius of the Mohr's circle to the distance to the Mohr-Coulomb criteria as shown in Fig. S5 (a). As the material is initially intact everywhere in the medium, the CF is thus smaller than 1 across the model. We chose a CF of 0.45 and the rest of parameters related with cohesion are derived to satisfy this condition. We then force a nucleation of the rupture by imposing a low peak strength patch around the area of nucleation. The exact location of the rupture initiation is arbitrary, the goal being to ensure unilateral propagation on the targeted fault. The length of this patch is greater than the nucleation length $L_c$. Fig. S5 (b) shows the distribution of the initial shear traction on the prescribed fault normalized by the peak strength, $\tau^0/\tau^p$. The grid size, $ds$, along the prescribed fault is set at 50 m. In this way, the number of finite elements in the estimated process zone size is assured to be between 8 and 14 on entire fault system.

**Acknowledgments**

Numerical modelling is performed using the High Performance Computing resources provided by the Institutional Computing program at Los Alamos National Laboratory. Imagery from CEOS_seismic pilot from ESA and ISIS CNES program, processed on S-CAPAD IPGP facility. E. Rougier & H. S. Bhat are grateful to P. Johnson (LANL) for initiating a collaboration between ENS and LANL. This work was partly funded by IRSN.

**Author contributions**

Y.K. and H.B. conceived the original idea. A.V., J.C., A.D. and Y.K. processed and analysed the optical data. K.O., E.R., Z.L., E.K., A.M. and H.B. designed the numerical tools and run simulations. R.L., S.B. and Y.K. analysed field data. Y.K., K.O. and H.B. wrote the manuscript with inputs from all authors.

**Competing interests:**

The authors declare no competing interests.

**Figures**

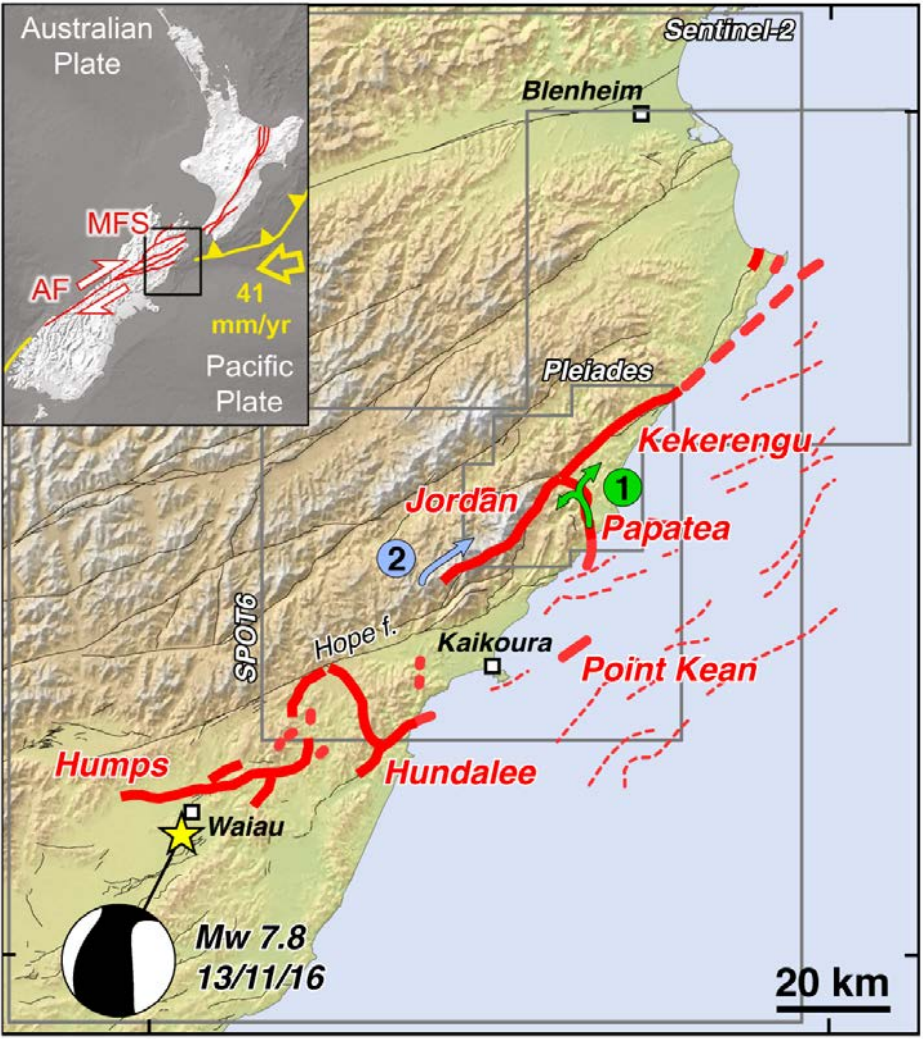

**Fig. 1. Map of the Kaikoura surface rupture** [11]. Footprint of satellite images is indicated. Labels 1 and 2 refer to alternative rupture scenarios.

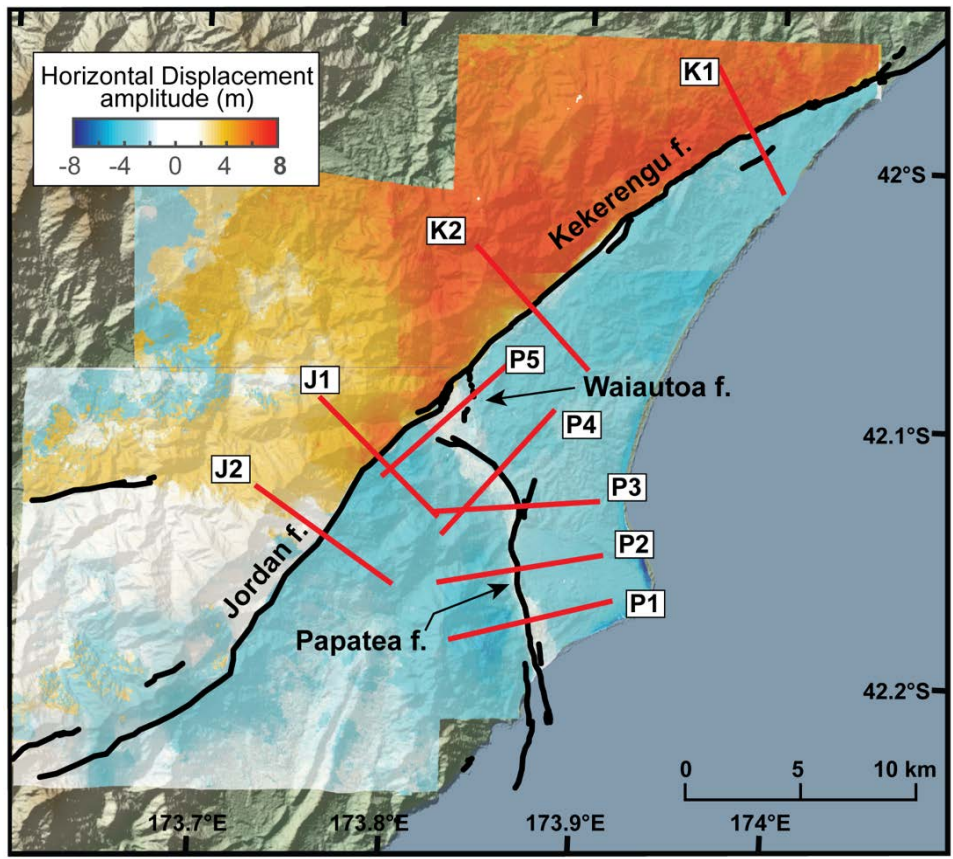

**Fig. 2. Deformation field around the triple junction.** Earthquake surface ruptures are indicated in black[15]. Color corresponds to the amplitude of horizontal displacement (positive towards North east). Low color saturation along the Papatea fault indicates off-fault damage. Red lines show the position of displacement profiles (See also Fig. S1).

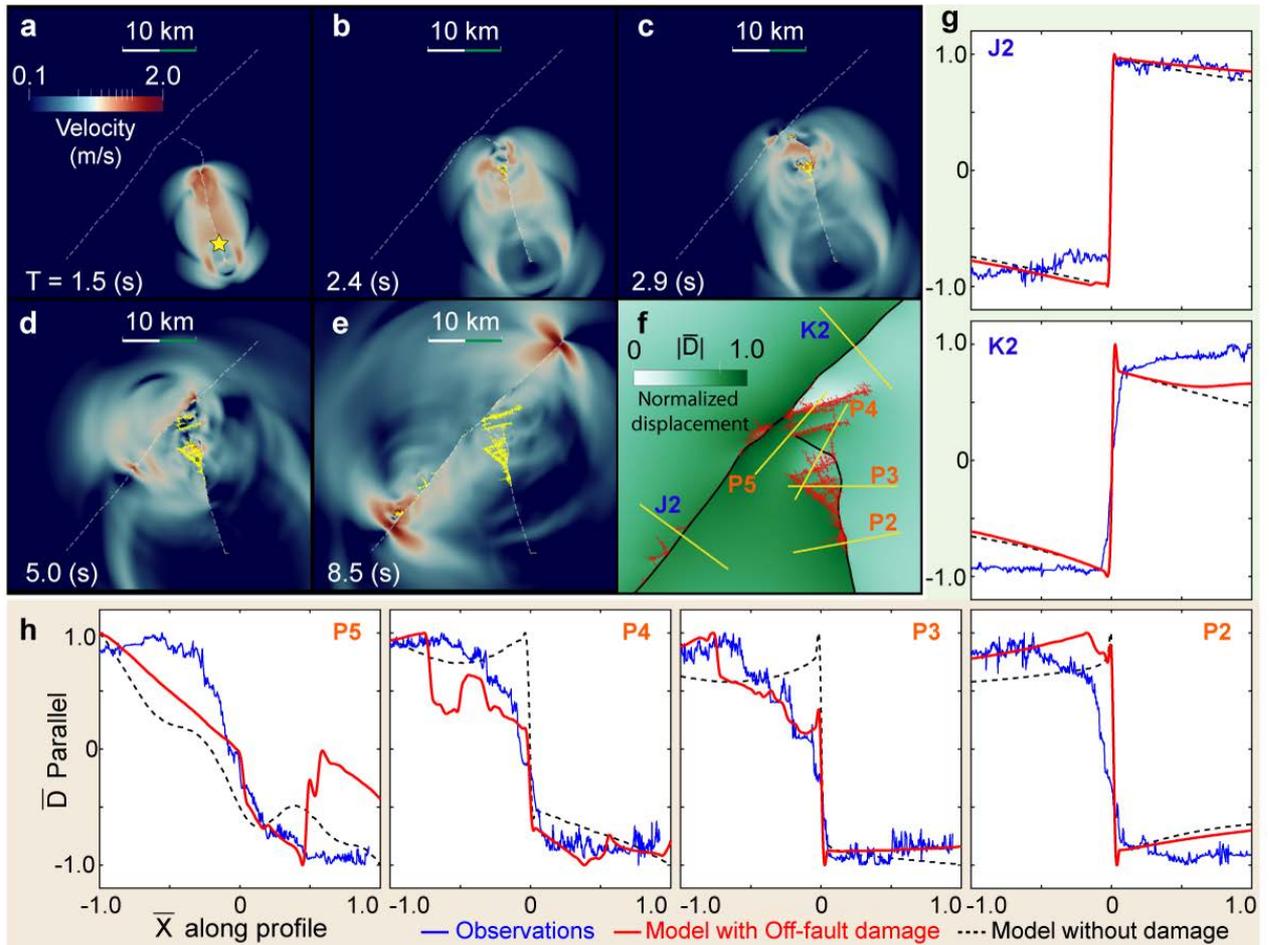

**Fig. 3. Rupture process, displacement field and profiles of fault parallel displacement for the first scenario.** (a-e) Snapshots of the velocity field associated with rupture nucleated on Papatea fault (yellow star). Dotted lines show the prescribed faults and yellow lines show the spontaneously activated off-fault crack network. (f) Deformation field and crack network at the end of earthquake event. (g, h) Displacement parallel to the fault along the different profiles across the prescribed faults. To focus on the broad features of the displacement field, both fault parallel displacement and distance along profile are scaled by their maximum values.

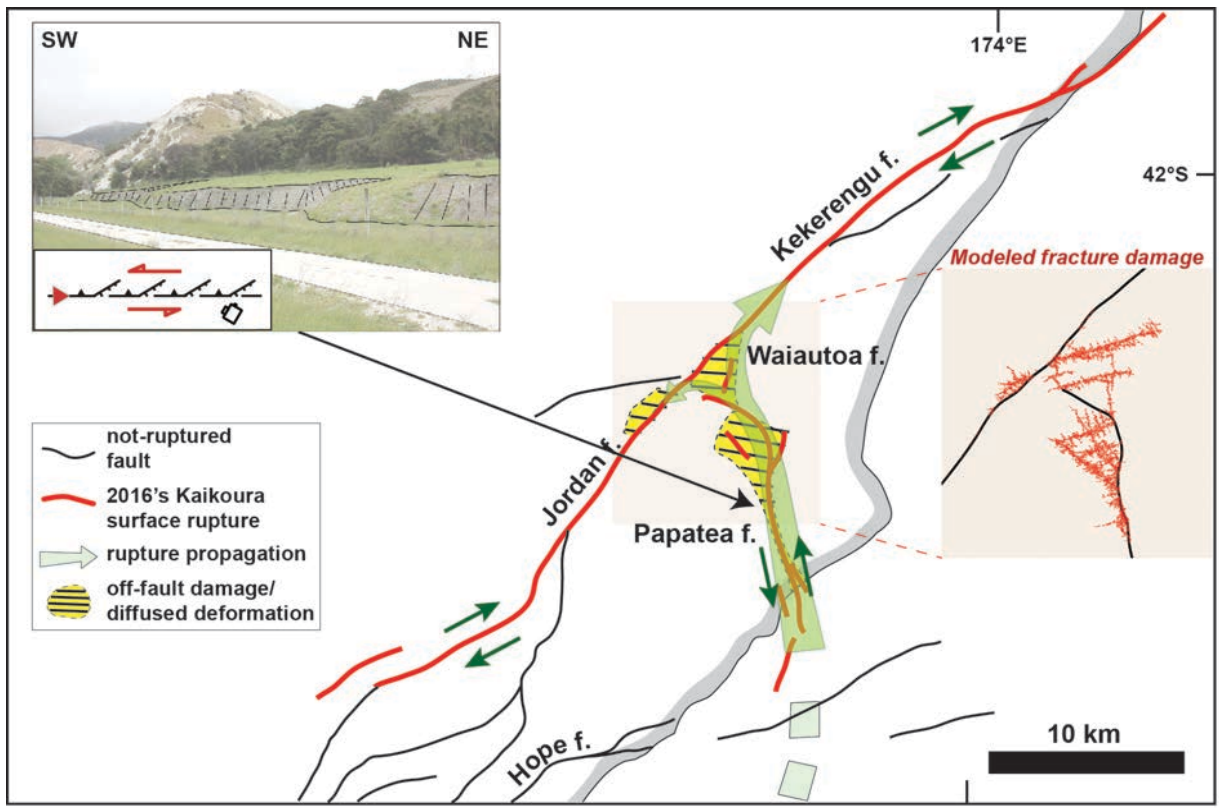

**Fig. 4. Summary of the preferred rupture path and associated fracture damage:** Photo inset shows the observed surface damage on the Papatea fault (S42°08'47'', E173°52'01''). In addition to the main strike-slip scarp with thrust, systematic westward branching with normal motion can be seen, which is best explained by left-lateral rupture propagating from the south.

# Earthquake damage patterns resolve complex rupture processes

**Authors:** Yann Klinger[1]*, Kurama Okubo[1,2], Amaury Vallage[1], Johann Champenois[1,3], Arthur Delorme[1], Esteban Rougier[4], Zhou Lei[4], Earl E. Knight[4], Antonio Munjiza[5], Stephane Baize[3], Robert Langridge[6] and Harsha S. Bhat[2]

**Affiliations :**
[1]Institut de Physique du Globe de Paris, Sorbonne Paris Cité, Université Paris Diderot, CNRS, Paris, France.
[2]Laboratoire de Géologie, École Normale Supérieure/CNRS UMR8538, PSL Research University, Paris 75005, France
[3]Seismic Hazard Division, Institut de Radioprotection et de Sûreté Nucléaire, Fontenay-aux-Roses, France
[4]EES-17 – Earth and Environmental Sciences Division, Los Alamos National Laboratory, NM, USA
[5]FGAG– University of Split, 21000 Split, Croatia.
[6]GNS Science, P.O. Box 30-368, Lower Hutt, 5040, New Zealand.

*Corresponding author. Email: klinger@ipgp.fr.

## Supplementary information

Supplementary information include
- A description of the second rupture scenario where rupture would start along the Jordan fault.
- Details about changes of state of stress, slip, and slip-velocity for scenario 1.
- A set of complementary figures:
  - Deformation field and displacement profiles across faults.
  - Correlation of Sentinel-2 images.
  - Slip-distribution along the different faults.
  - Model description and schematic of the FDEM.
  - Definition of the closeness to failure and initial shear traction.
  - Rupture process, displacement field and profiles for scenario 2.
  - Secondary crack network due to dynamic earthquake propagation along with change of stress, slip and slip-velocity in function of time during rupture propagation.
- A table of physical parameters used in modeling.
- Caption for the two movies showing rupture propagation for scenario 1 and 2.

**Second scenario: rupture nucleation at the southern end of the Jordan thrust**

The model parameters used in this simulation is exactly same with the first scenario discussed in (Fig. 3). Fig. S6 (a) to (e) show the snapshots of the second scenario. In this scenario the rupture propagates northward and activates off-fault cracks. We found a small nucleation of the rupture at the main kink of the Papatea fault as shown in Fig. S6 (c), which then propagates bilaterally. The rupture propagating southward is in fact trapped as shown in Fig. S6 (e) due to the kink, creating new cracks east of the Papatea fault. The rupture along the Kekerengu fault accelerated fast enough to transition to supershear speeds. The pre-stress state is partially preferable for a transition to a supershear rupture due to the fault geometry. The nucleation of a daughter crack is clearly seen in Fig. S6 (c), propagating northward on the Kekerengu fault. Fig. S6 (f) shows the damage pattern and the displacement field at the end of the simulation, where all particle motion ceases. Since the rupture is arrested at the north of the Papatea fault, slip is not observed on the southern part of this fault. Fig. S6 (g) and (h) show the profiles on the Jordan thrust-Kekerengu fault and the Papatea fault respectively. The model is still compatible with the observations on (g), whereas it barely fits with observations even with off-fault damage because there is no significant damage to the west of the Papatea fault. Furthermore, the localized slip is no longer observed with off-fault damage in Fig. S6 (h) on profile P2. The deformation in this case simply reflects the large slip on the Jordan thrust – The Kekerengu fault. We therefore conclude that this scenario is less likely than the first scenario.

**Stress change, slip and slip velocity on the Jordan, the Kekerengu and the Papatea faults for the first scenario**

We computed the mechanical fields on the two faults separately as shown in Fig. S7. Fig. S7 (a) shows the trace of the Papatea fault and the dynamically activated off-fault cracks plotted in red. Although it forms an intricate crack network around the main kink of the fault, we find a large chain of cracks in the direction towards northwest, which plays a role in the distributed displacement profiles. As the Papatea fault has relatively large kinks and the initial normal and shear tractions on the fault are therefore heterogeneous, the change of normal stress and stress drop along the fault is significant as shown in Fig. S7 (b), (c) and (d). The comparison between the model with off-fault damage (in red) and the purely elastic model (in blue) of the change of normal stress indicates that the off-fault medium cannot sustain large stress concentrations as shown at $x/L = 0.72$ in Fig. S7 (c). We also find a locally negative stress drop around $x/L = 0.72$, where the angle of maximum compressional principal stress is fairly orthogonal and thus the initial shear traction is relatively small. Hence, this part can cause negative stress drop after rupture propagation on such a non-planar fault. Fig. S7 (e) shows the accumulated slip distribution on the Papatea fault. We found a locally enhanced slip in the case with off-fault damage at $x/L = 0.62$ in Fig. S7 (e), which is directly induced by the off-fault cracks in the vicinity of the fault. Fig. S7 (f) shows the slip velocity in time and space, which shows the detailed rupture process on both faults. The rupture is initially nucleated around $x/L = 0.3$, propagating bilaterally on the Papatea fault. When the rupture reaches $x/L = 0.7$, it arrests and immediately jumps ahead at $x/L = 0.83$, propagating bilaterally. Eventually the whole length of the Papatea fault is ruptured in this scenario. The slip velocity is remarkably perturbed by the spontaneous off-fault cracking. Since the stress distribution is extremely perturbed by the crack network, negative slip velocity is temporarily induced around $x/L = 0.62$ at $t = 6$ s. Fig. S7 (g) to (l) shows the same quantities on the Kekerengu fault. As it has less geometrical complexity compared to the Papatea fault, there is less off-fault damage on the Kekerengu fault as shown in Fig. S7 (g), which leads to the well-localized slip as shown in Fig. S7 (g). The change of normal stress is also smoothed by the off-fault damage as shown in Fig. S7 (i).

**Supplementary Figures**

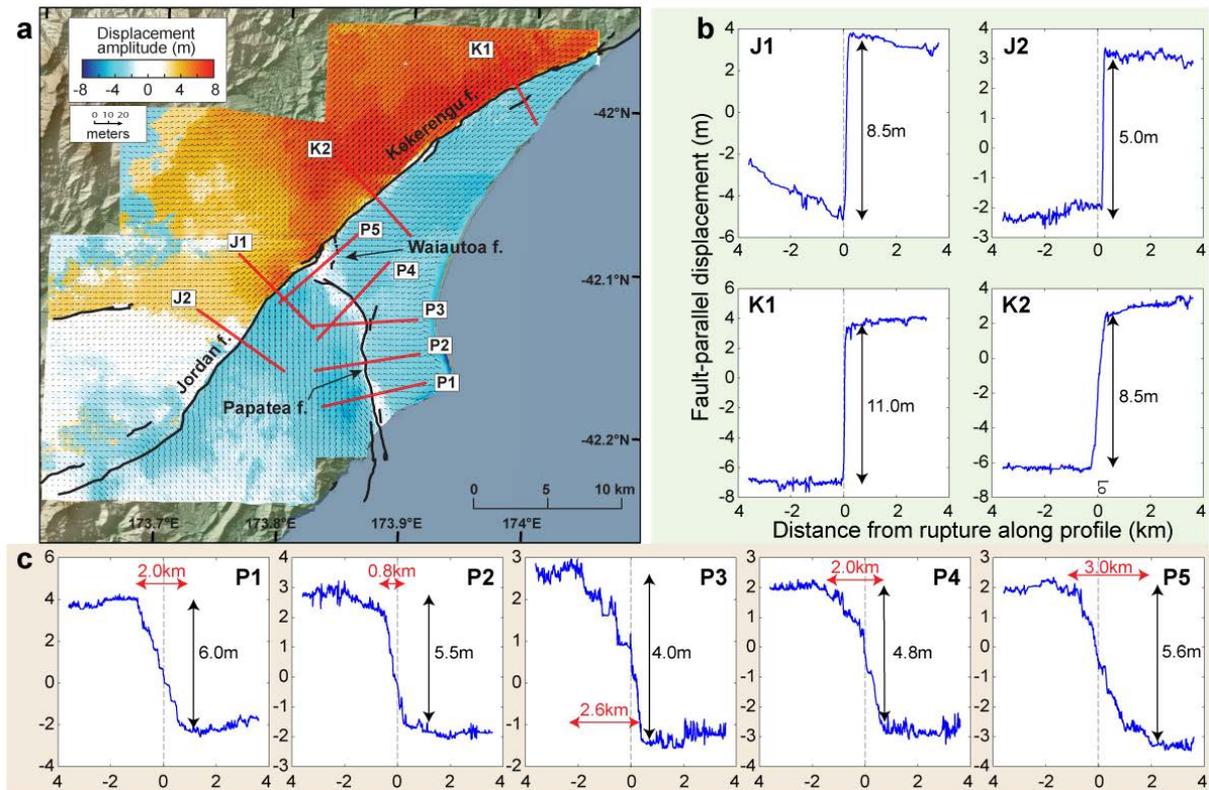

**Fig. S1. Deformation field and associated displacement profiles.** (a) Deformation field around the triple junction with the azimuth of horizontal displacement (arrows). Surface ruptures related to the Kaikoura event are in black[15]. The size of the arrow scales with the amplitude. Arrows converging or diverging from the fault indicate respectively some component of thrust or normal motion. The Papatea block is slightly rotating counter-clock wise. (b) Profiles on the Jordan thrust (J1 and J2) and the Kekerengu faults (K1 and K2). (c) Profiles on the Papatea fault (P1 to P5).

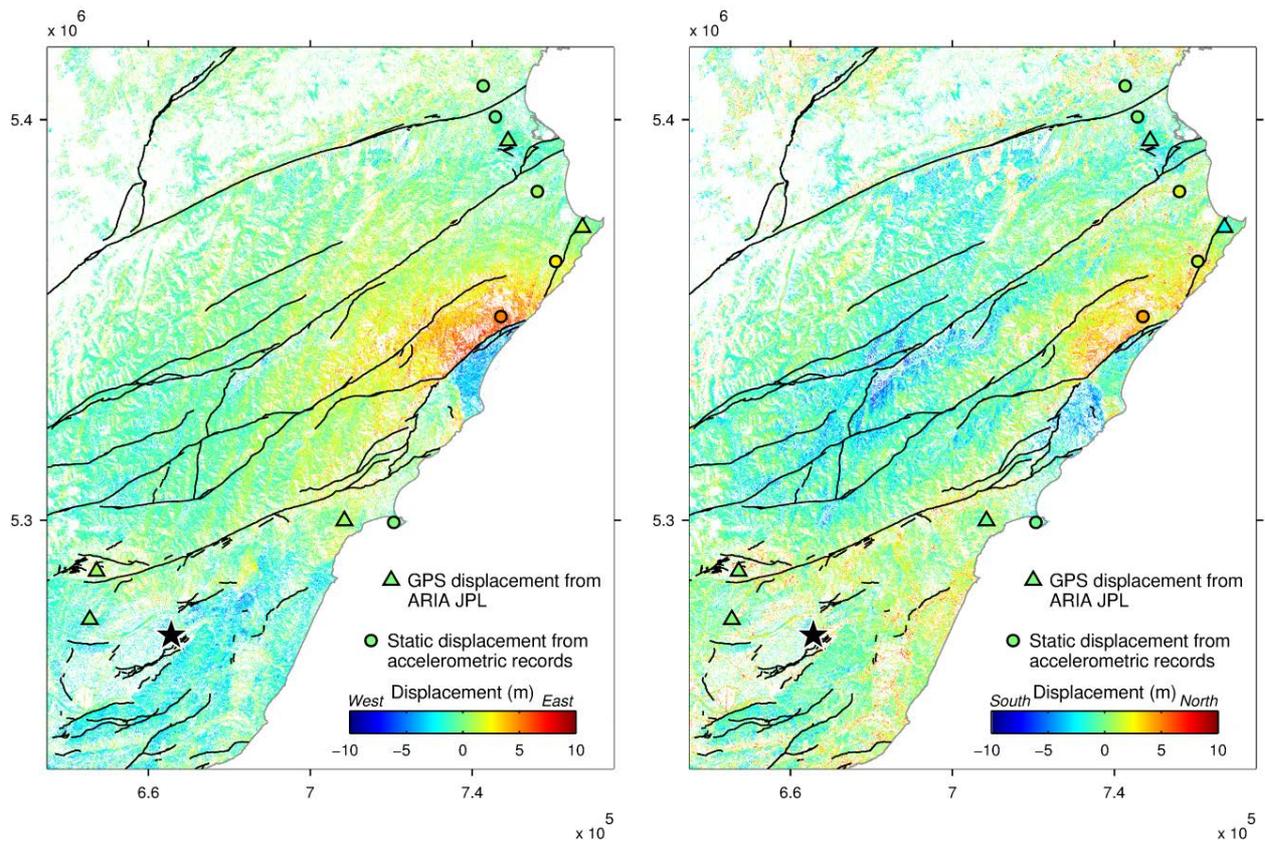

**Fig. S2. East-West and North-South components of displacement computed from the correlation of Sentinel-2 images.** Coordinates are in UTM. Consistency of the results is checked by comparison with GPS and static motion derived from local strong-motion instruments, for each component. The far-field displacement is set to be zero. The fault network (black lines) is from GNS (https://data.gns.cri.nz/af/).

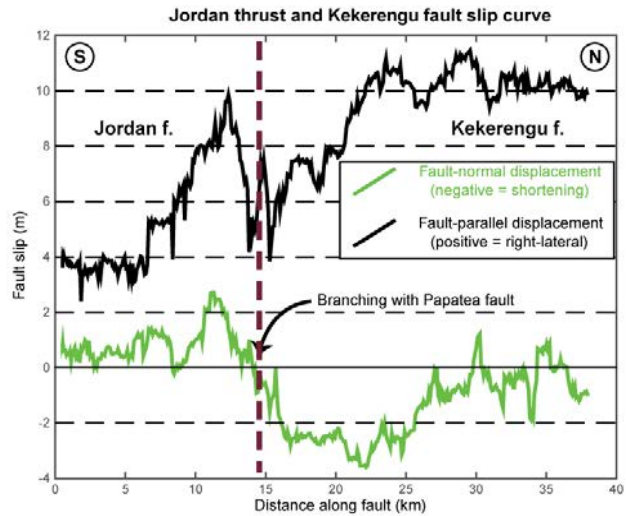

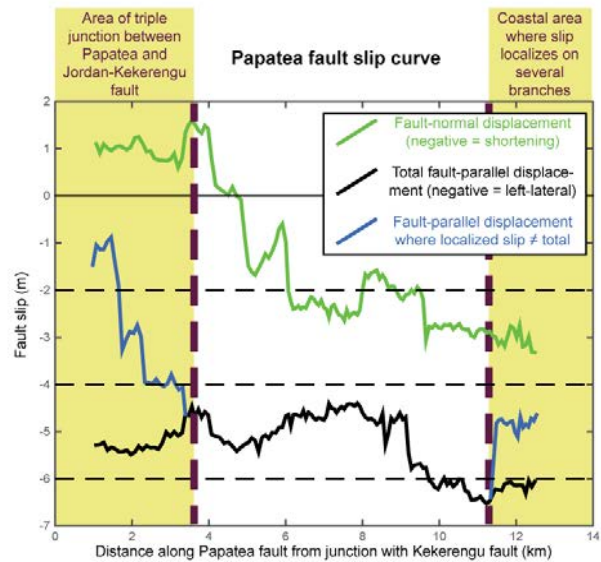

**Fig. S3. Slip distribution for the two components of horizontal motion, parallel and perpendicular to the fault, for the Kekerengu – Jordan fault system, and for the Papatea fault.** The slip is measured every 90 m, using 8 km-long and 90m wide swath, with no overlap between successive swaths. The general shape of the slip distribution is consistent with lower resolution slip distribution[12], although details of slip variation can be seen that correspond to variation in fault geometry. The thrust component of the Kekerengu fault and the normal component of the Jordan fault are clearly visible. The thrust component of the Papatea fault is visible almost all along the fault section. The two yellow end boxes indicate locations where the amount of off-fault damage is large.

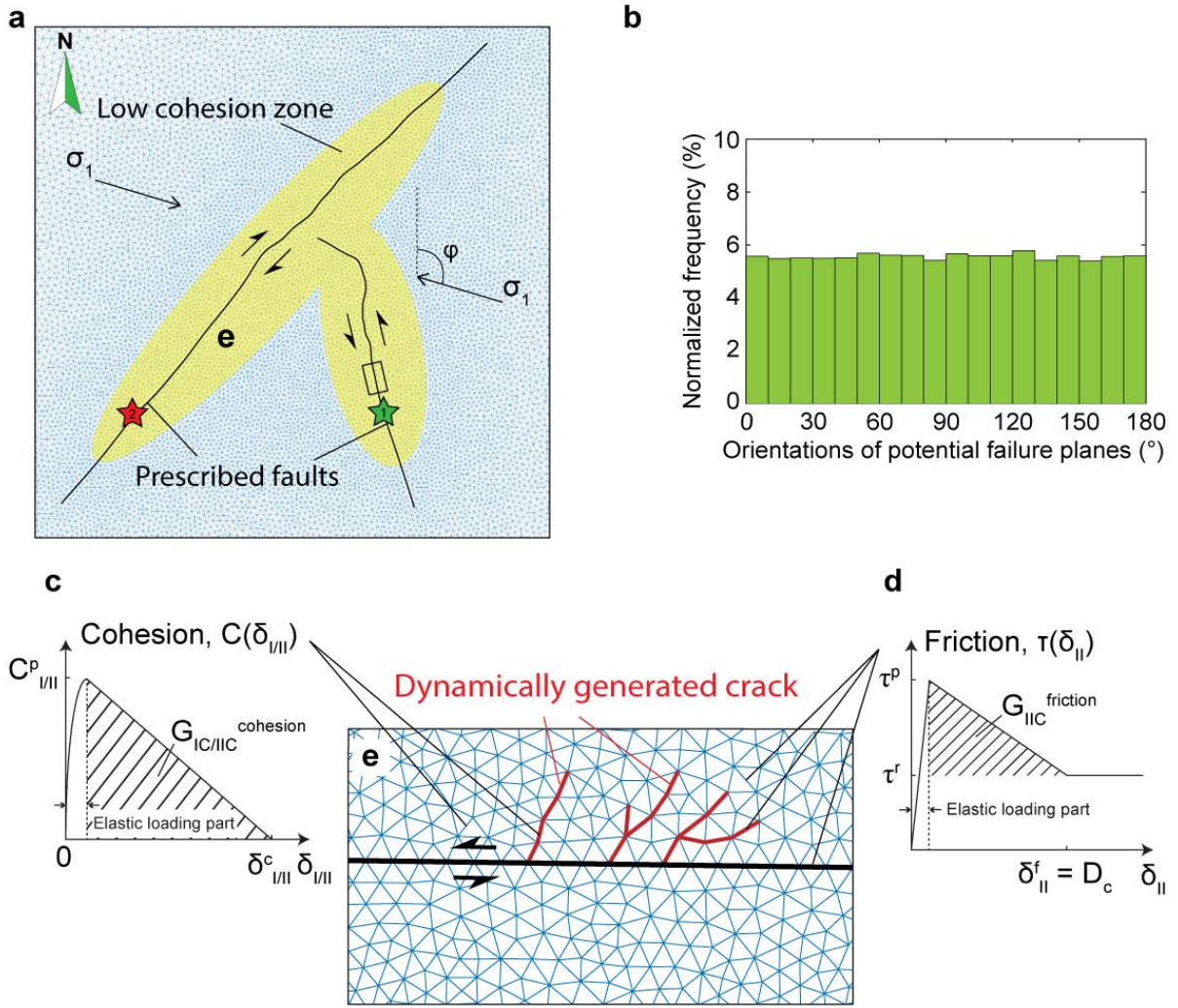

**Fig. S4. Model description and schematics of FDEM.** (a) Schematics of mesh discretization on the prescribed fault system. The Jordan fault, the Kekerengu fault and the Papatea fault are traced as shown in solid black line. Blue lines show the discrete finite elements. The mesh size is exaggerated for clarity purposes. The overall domain size is 90km x 90km, while the prescribed faults are in 30km x 30km in the middle of the domain to avoid the effects of wave reflections from the domain boundaries. The total number of finite elements is 514,000. $\sigma_1$ is the maximum compressional principal stress and $\varphi$ is the angle of $\sigma_1$ to the north. Arrows show the sense of slip. The areas of weakened material are highlighted in yellow. Green and red stars show the position of the rupture nucleation for the first and second scenarios

respectively. Small box shows the zoomed window shown in (e). (b) Histogram of the orientations of the potential failure planes. (c) Linear softening cohesion curve. $\delta_{I/II}$ is the amount of slip in tensile (mode I) and shear (mode II). $C^p_{I/II}$ is the peak cohesive strength in tension and shear. $\delta^c_{I/II}$ is the critical normal/tangential displacement for softening of tensile/shear cohesion. $G_{IC/IIC}^{cohesion}$ is the dissipated energy by breaking cohesion. (d) Linear slip-weakening curve. $\delta^f_{II}$ is the characteristic slip distance which is identical with the $D_c$ in conventional slip-weakening law. $\tau^p$ and $\tau^r$ are the peak and residual strength in friction, derived as $\tau^p = f_s\sigma_n$ and $\tau^r = f_d\sigma_n$, where $\sigma_n$ is the compressive normal stress on the boundary of elements. $G_{IIC}^{friction}$ is the fracture energy dissipated by the frictional process. (e) Zoomed window around faults shown in (a). Black solid line shows the prescribed fault and blue lines show the finite elements. Red lines show the newly generated cracks on which the cohesion starts to brake due to the stress concentration by the dynamic rupture.

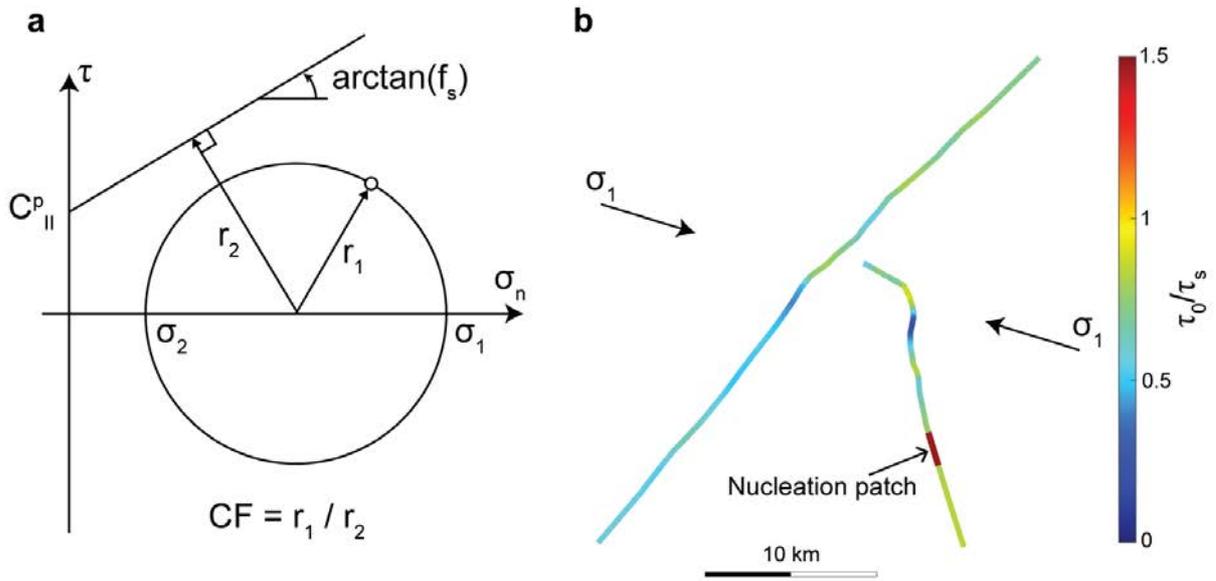

**Fig. S5. Definition of closeness to failure (CF) and the initial shear traction.** (a) Schematic of closeness to failure CF defined as $r_1/r_2$. (b) The distribution of initial shear traction normalized by the frictional strength and nucleation patch.

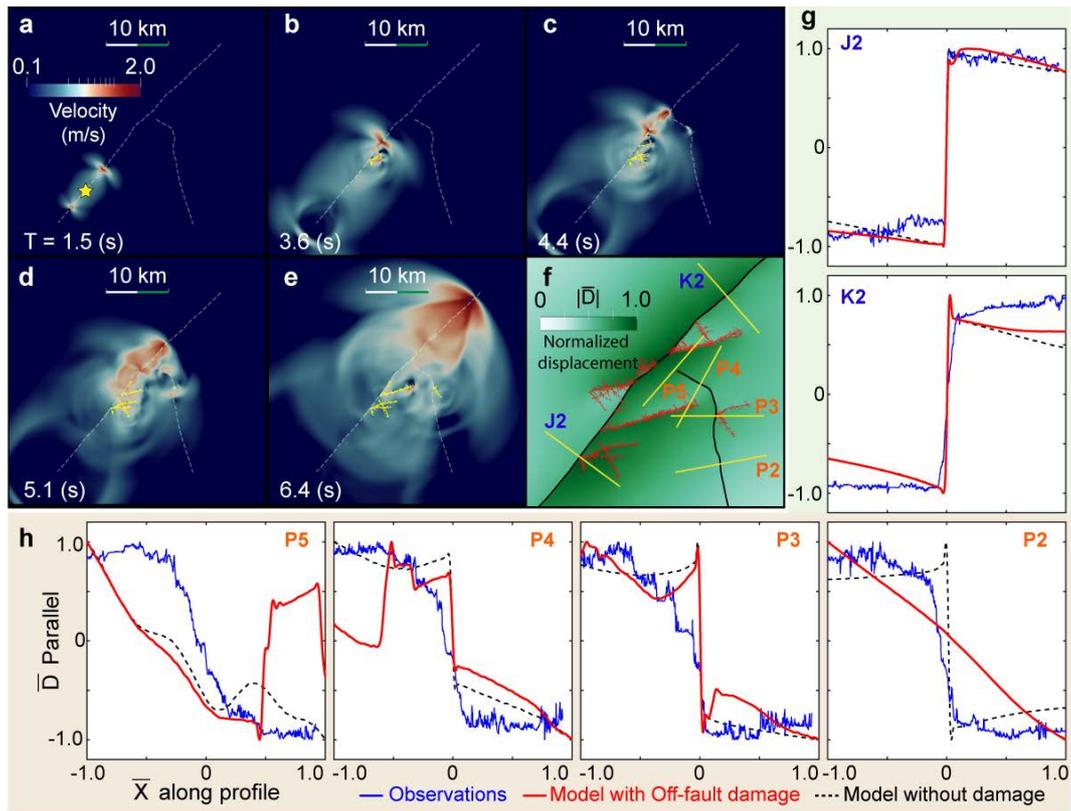

**Fig. S6. Rupture process, displacement field and profiles of displacement parallel to the fault for the second scenario.** The rupture is nucleated at the southern end of the Jordan fault (yellow star). (a to e) Snapshots of rupture from south of the Papatea fault to the Kekerengu fault. Dotted line shows pre-existing faults and yellow lines show the secondary crack network generated by the dynamic earthquake rupture propagation on the main faults. The color contours show the particle velocity magnitude. (f) The displacement field and the crack network obtained at the end of the earthquake event (at 18s). The yellow lines across the main faults show the position of profiles, i.e., the profile of the displacements parallel to the fault shown in (g) and (h). The blue line shows the observations and the red line shows the model results with off-fault damage. The dotted black line shows the model results when considering a purely elastic medium, which does not allow for off-fault damage. Both the displacement parallel to the fault and the distance from the rupture are normalized with the range of displacement and the length of the profiles respectively.

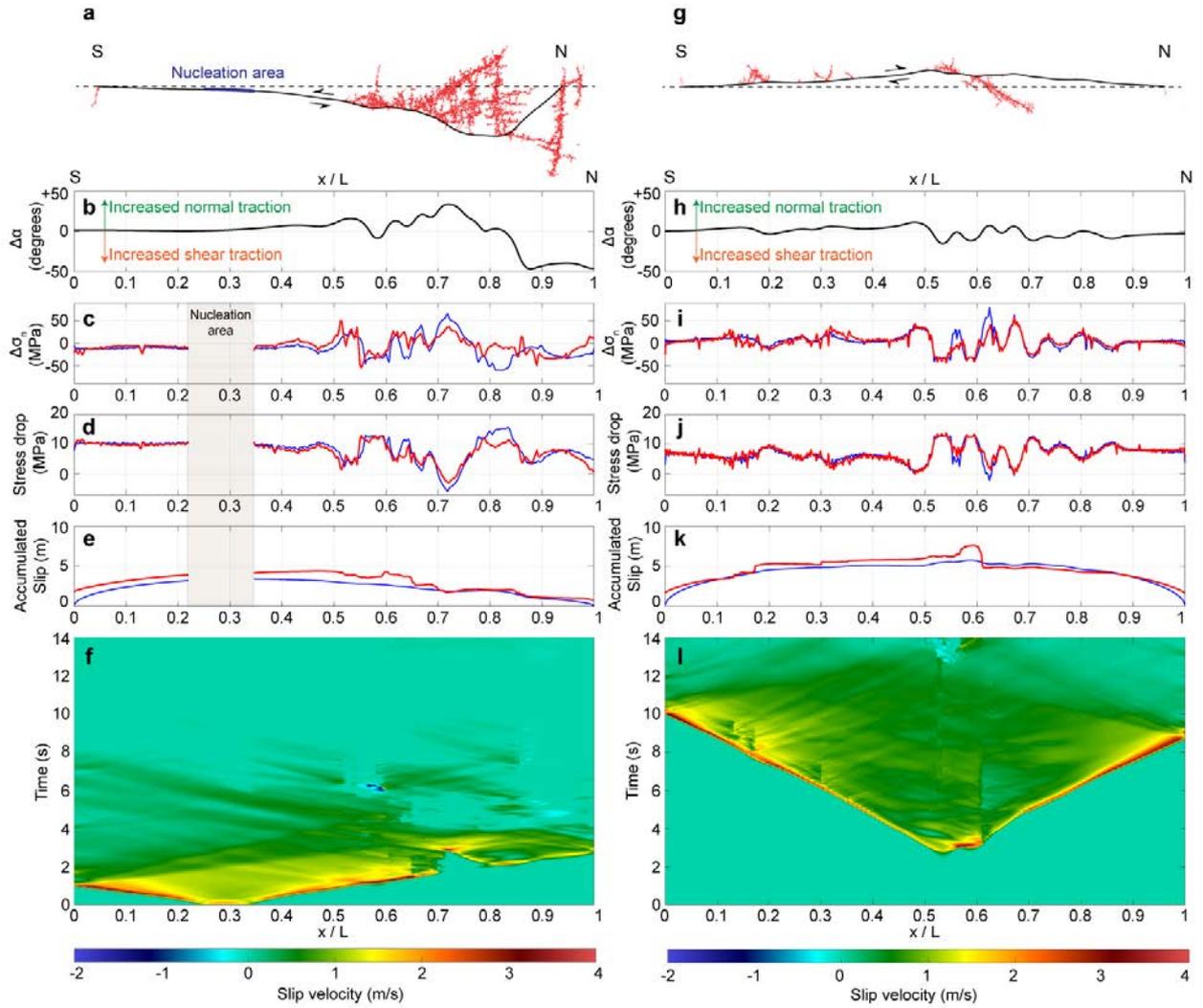

**Fig. S7.** (a) The trace of secondary crack network generated/activated by dynamic earthquake rupture (in red) on the Papatea fault. The rupture is artificially nucleated at the nucleation segment indicated in blue. Notation S (south) and N (north) show the direction of the fault. The dotted auxiliary line shows a reference to measure the angle of the maximum compressional principal stress, $\Delta\alpha$, to the fault shown in (b). $\Delta\alpha$ indirectly indicates the ratio of normal traction to shear traction. Positive values of $\Delta\alpha$ indicate larger normal traction than the reference traction state on the auxiliary line, whereas negative values show smaller ratio of the normal traction to the shear traction. The angle of maximum compressional principal stress to the reference is 53.6° on the Papatea fault and 64.8° on the Kekerengu fault. (c), (d), (e) and (f) show the change of normal stress $\sigma_n^0 - \sigma_n^1$, stress drop $\tau^0 - \tau^1$, accumulated slip and slip velocity respectively. The red line shows the model with off-fault damage and blue shows the model without off-fault cracks. The color contours show the evolution of the slip velocity on the fault. The horizontal axis shows the position normalized by the length of fault, x/L = 0 corresponding to the southern edge of the fault. (g), (h), (i), (j), (k) and (l) show the same quantities on the Kekerengu fault.

| Variables | Values | Description |
|---|---|---|
| $\rho$ | 2700 kg/m³ | Density |
| $E$ | 75 GPa | Young's modulus |
| $\mu$ | 30 GPa | Shear modulus |
| $\nu$ | 0.25 | Poisson's ratio |
| $\sigma_1$ | 45.4 MPa | Maximum compressional principal stress |
| $\sigma_2$ | 19.1 MPa | Minimum compressional principal stress |
| $\phi$ | 107 ° | Angle of σ₁ to the north |
| $ds$ | 50 m | Grid size on fault |
| **On prescribed fault** | | |
| $f_s$ | 0.4 | Static friction coefficient |
| $f_d$ | 0.1 | Dynamic friction coefficient |
| $\delta_{II} = D_c$ | 0.17 m | Characteristic slip distance |
| **On off-fault medium** | | |
| $f_s$ | 0.5 | Static friction coefficient |
| $f_d$ | 0.15 | Dynamic friction coefficient |
| $\delta_{II} = D_c$ | 0.017 m | Characteristic slip distance |
| $C^p_I$ | 8 MPa / 30 MPa | Peak cohesion for mode I opening crack (Low cohesion zone/the rest of domain) |
| $C^p_{II}$ | 30 MPa / 100 MPa | Peak cohesion for mode II shear crack (Low cohesion zone/the rest of domain) |
| $\delta^c_I$ | 2.7 mm | Critical normal displacement for softening of tensile cohesion |
| $\delta^c_{II}$ | 7.5 mm | Critical tangential displacement for softening of shear cohesion |

**Table S1.** Parameters used in numerical modelling

**Legends of supplementary movies**

**Movie MS1.**

Particle velocity field, slip rate and acceleration records for rupture nucleation on the Papatea fault. The yellow lines correspond to spontaneously activated off-fault fractures.

**Movie MS2.**

Particle velocity field, slip rate and acceleration records for rupture nucleation on the Jordan fault. The yellow lines correspond to spontaneously activated off-fault fractures.